\documentclass[12pt]{article}
\usepackage{amsfonts}
\newcommand{\p}[1]{(\ref{#1})}
\newcommand{\cp}{\mbox{$\cal P$}}
\newcommand{\e}{\eta}
\newcommand{\be}{\begin{equation}}
\newcommand{\bea}{\begin{eqnarray}}
\newcommand{\ee}{\end{equation}}
\newcommand{\eea}{\end{eqnarray}}
\newcommand{\la}[1]{\langle S_{#1}| }
\newcommand{\ra}[1]{|S_{#1}\rangle }

\def\theequation{\arabic{section}.\arabic{equation}}
\begin{document}
\setcounter{page}0
\renewcommand{\thefootnote}{\fnsymbol{footnote}}
\thispagestyle{empty}

\vspace{1.5cm}

\begin{center}
{\large\bf
Lagrangian formulation of the massless higher integer spin fields
 in the AdS background
}\vspace{0.5cm} \\

I.L. Buchbinder${}^{a}$\footnote{E-mail:
joseph@tspu.edu.ru}, A. Pashnev${}^{b}$\footnote{E-mail:
pashnev@thsun1.jinr.dubna.su}
and
M. Tsulaia${}^{b,c}$\footnote{E-mail:
tsulaia@thsun1.jinr.ru}\\
\vspace{1.5cm}

${}^a${\it Department Theoretical Physics,
Tomsk State Pedagogical University}\\
{\it Tomsk, 634041, Russia}\\
\vspace{0.5cm}

${}^b${\it Bogoliubov Laboratory of Theoretical Physics, JINR} \\
{\it Dubna, 141980, Russia}\\
~\\
${}^c${\it The Andronikashvili Institute of Physics, Georgian Academy
of Sciences,}\\
{\it Tbilisi, 380077, Georgia} \vspace{1.5cm}\\

{\bf Abstract}
\end{center}
\vspace{1cm}

We construct the Lagrangian description of  arbitrary integer
higher spin massless fields on the background of the D -- dimensional
anti -- de Sitter space. The operator constraints in auxiliary Fock
space corresponding to subsidiary conditions for irreducible unitary
massless representations of the D -- dimensional anti -- de Sitter group are
formulated. Unlike  flat space, the algebra of the constraints
turns out to be nonlinear and analogous to the $W_3$ algebra. We
construct the nilpotent BRST charge for this nonlinear algebra and
derive on its basis the correct field content and gauge invariant action
describing the consistent arbitrary integer spin field dynamics in
$AdS$ space.

\newpage\renewcommand{\thefootnote}{\arabic{footnote}}
\setcounter{footnote}0\setcounter{equation}0
\section{Introduction}

Construction of  interacting higher spin field models is one of
the most intriguing problems of field theory. The Lagrangian
formulation for free higher spin fields is known for a long time
although some aspects of such fields excite interest so far. A general
problem of  higher spin fields interacting with each other or
coupled to an arbitrary background is still open.

At present, one of the main stimulus to study  interacting higher
spin fields comes from the string theory. As is well known, any string model
contains an infinite tower of massive higher spin particles in its
spectrum (see e.g. \cite{GSW}) which, in principle, can be treated as
massless at the energies above the string scale.
The string interaction implies
 interactions of all these particles among themselves. As a result, an
interacting field theory, formally corresponding to some string model,
should in principle contain an infinite tower of higher spin fields as
well (see \cite{Vasiliev} for a review of some features of
interacting massless higher spin fields). The question that naturally
arises is whether one can construct a field theory  which describes
 higher spin fields interacting among themselves or with some
gravitational background and contain a finite number of fields at the
same time.  An example of such theories is given by field models
in an external gravitational field with constant curvature.

The higher spin field theories coupled to a constant
curvature gravitational background have recently attracted much
attention and their various aspects were investigated in
detail \cite{BGP} -- \cite{Z}. It has been pointed out that the fields
on the D-dimensional anti -- de Sitter ($AdS_D$) background
possess  remarkable properties. The $AdS$ manifold is compatible
with supersymmetry (see e.g. \cite{BK} for $D=4$), $AdS$ space is a
most natural vacuum where the interacting massless higher spin
fields can consistently propagate \cite{FV},   $AdS$
space also plays the important role in the modern
string/brane theory \cite{Maldacena}.

There exist two different approaches for the description of
higher spins both in the flat and in AdS spaces. The first
one uses the totally symmetric fields or fields with the another
 symmetry properly described by the corresponding Young tableaux.
For the time being the Lagrangians for totally symmetric
fields carrying the arbitrary spin $s$ were known only in
the $AdS_4$ space for nonsupersymmetric \cite{F1} and supersymmetric
\cite{KS} (see, e.g., \cite{BK} for the
description of supersymmetry on
$AdS_4$ space) theories.

The alternative approach, so called gauge approach, uses for the
description of higher spins the geometrical objects like vielbeins and
connections \cite{FOR} -- \cite{LV}, \cite{AD}.
This approach  formulated in
terms of differential forms
turned
out  to be very  efficient  also for the construction
of  the theory of free higher
spin fileds in $AdS_4$ \cite{FOR}  and in $AdS_D$ \cite{LV} spaces.
Both  approaches mentioned above
are equivalent to each other. However, in contrast with previous one,
 the gauge approach includes a number of
 auxiliary fields having the geometrical origin. These auxiliary fields
  should be eliminated in order to
obtain the final Lagrangians in terms of totally symmetric fields.
 In the gauge approach however
the final Lagrangians
for the theories of the higher spin fileds on $AdS_D$
 have not been
constructed in explicit form though,
the Lagrangians containing all (still non -- eliminated) geometrical
auxiliary fields were given.

In a recent paper \cite{Z}, the Lagrangian
description
of the massive higher spin fields propagating through
the $AdS_D$ background was also obtained.  However, the corresponding
action for the massless higher spin fields cannot be directly derived
from  the action for  massive higher spin fields simply setting the
mass parameter $m$ equal to zero.  The reason is that in general the
structure of off -- shell constraints on
the basic fields in a massless
theory  differs from  a massive one and cannot be obtained at the
formal limit $m=0$ in a massive theory.

For
the fields belonging to  arbitrary
massless irreducible representations of $AdS_D$
the completely self -- consistent
equations of motion and gauge transformations were formulated
in \cite{Me} --
\cite{BMV}.  Therefore it seems tempting to explicitly construct
the corresponding {\em Lagrangians} from which the above equations of
motion follow.

In the present paper, we develop the method of the BRST-construction,
earlier used in \cite{OS} -- \cite{BPT2} for the description of
free higher spin fields in Minkowski space-time (see also \cite{LA1}
for  different treatments and \cite{S} for the Euclidean version),
to derive the Lagrangians describing the
interaction of  massless arbitrary higher spin $s$ fields with the
$AdS_D$ background  which has also an arbitrary number $D$ of space --
time dimensions.  The advantage of this method is that it automatically
leads to  the Lagrangians which possess necessary gauge invariance
required to remove  nonphysical states -- the ghost from the
spectrum.  Moreover, if one excludes all auxiliary fields using the BRST
gauge invariance and thus completely fixes the BRST gauge, the field
which is left with will obey the correct physical conditions.  We
 show that the  BRST approach, developed for the Lagrangian description
 of various representations of the Poincare group in the flat
space -- time, admits the corresponding deformation to  $AdS_D$
spaces as well. Below, we consider massless representations of the
$AdS_D$ group when the corresponding Young tableaux describing the
irreducible representations of the flat space massless little group
$O(D-2)$ has only one row.  For this  kind of representations, the
effect observed in \cite{BMV}, of decomposition of irreducible
representations of the $AdS_D$ group into  several irreducible
representations of $O(D-2)$, when taking the flat limit, does not take
place and we are left with only one physical field in both  flat and
curved space -- time.  As it could be expected, the resulting Lagrangian
is simultaneously the $AdS_D$ generalization of the flat -- space
Lagrangian for the massless irreducible higher spin fields \cite{F} as
well as $D$ -- dimensional generalization of the $AdS_4$ Lagrangian of
\cite{F1}.

One should note the paper \cite{B2} in which the BRST construction was
applied to the description of massless higher spins in $AdS_D$ space --time.
Our approach differs from the one given in this paper in some points.
First of all, we describe  $AdS_D$ space -- time as  space parameterized
by $D$ intrinsic coordinates with the metric of constant curvature. Such
a consideration greatly simplifies all construction in comparison with
the embedding of  $AdS_D$ space in  $D+1$ flat space with the
help of embedding constraints \cite {B2} which should be verified at
some steps in addition to the general procedure. Besides, from the very
beginning we include in the BRST charge the constraints responsible for
the tracelessness of the basic tensor field. The resulting BRST charge
is nilpotent and provides the possibility of deriving step by step a
final gauge
invariant action for the basic field on the basis of the
universal general procedure.

The paper is organized as follows.

In Sec 2., we derive  nilpotent BRST charge for the system of constraints
which include the generalized d'Alambertian, transversality and
tracelessness operators
in the $AdS_D$ space -- time.

In Sec 3., we explicitly show that after the
 partial BRST gauge fixing one is left with only
one physical field which correctly describes the
 massless irreducible higher spin fields in the
$AdS_D$ background and present the corresponding gauge
invariant Lagrangian.

The procedure of the
partial BRST gauge fixing is carried out in Appendix.

\setcounter{equation}0\section{System of constraints and BRST charge}

In order to construct  field theoretical Lagrangians describing
the interaction of  massless higher spin fields with the $AdS_D$
background, it is convenient to introduce the auxiliary Fock space
 spanned
by the creation and annihilation operators $a^+_\alpha$ and $a_\alpha$
\be \label{osc}
\left[ a_\alpha,a_\beta^+ \right] =\eta_{\alpha \beta},\;\quad
\eta_{\alpha \beta}=diag(-1,1,1,...,1),
\ee
where $\alpha$ and $\beta$
are the tangent space indices ($\alpha , \beta = 0,1,..., D-1$), raised
and lowered by the tangent space metric $\eta_{\alpha \beta}$ and its inverse
$\eta^{\alpha \beta}$ in the usual way.

An arbitrary vector in this Fock space thus has the form
\be \label{Fockvector}
|\Phi\rangle =\sum_n
\Phi_{\alpha_1\alpha_2\cdots\alpha_{n}}
(x)
a^{\alpha_1 +}a^{\alpha_2 +}\cdots a^{\alpha_n +}
|0\rangle.\nonumber
\ee
Obviously  fields $\Phi_{\alpha_1\alpha_2\cdots\alpha_{n}}(x)$
are automatically symmetric with respect to the permutation of their
indices,  therefore,
representations of the corresponding flat space little group $O(d-2)$
are characterized by the Young tableaux with one row.

The irreducible massless representations of the $AdS_D$ group are
realized in the space of totally symmetric, transversal and traceless
fields \cite{Me}. Hence, in order to provide these conditions we
have to impose some subsidiary constraints on the vectors \p{Fockvector}.

The  transversality condition on the  basic field $|\Phi\rangle$
is taken in the form
\be \label{nogh}
L_1|\Phi\rangle= a^\mu p_\mu|\Phi\rangle=0
\ee
where
$a_\mu^+ = e^\alpha_\mu a^+_\alpha,
\quad a_\mu = e^\alpha_\mu a_\alpha,$
 the operator $p_\mu$ is the covariant derivative \cite{KL}-\cite{Me}
\be \nonumber
p_\mu = \partial_\mu + \omega_\mu{}^{\alpha \beta}a^+_\alpha a_\beta
\ee
while $e^\alpha_\mu$ and $\omega_\mu{}^{\alpha \beta}$ are the
 vielbein and spin connection, respectively.

The tracelessness condition of the field
$|\Phi\rangle$
can be expressed in terms of the operator $L_2 = \frac{1}{2}a^\mu
a_\mu$ as
\be \label{notr} L_2|\Phi\rangle=0 \ee

Our goal is to explicitly construct the Lagrangian from which
the tracelessness, transversality  and the mass -- shell
conditions follow as a consequence of the equations of motion.
We have  already mentioned in
the introduction that the method of the BRST constructions seems to be most
appropriate and straightforward for these purposes.

To obtain
a complete set of constraints required for the construction of the
nilpotent hermitian BRST -charge, we introduce the operators
$L^+_1= a^{\mu +}p_\mu$, $L^{+}_2= \frac{1}{2}a^{\mu +}a_{\mu}^+$ which
are conjugate to the operators $L_1$ and $L_2$ with respect to the
invariant integration measure $d^Dx \sqrt{-g}$ as well as
the covariant  d'Alambertian\footnote{In order to justify the
inclusion of a mass -shell operator
in the form \p{lapl}
into the total set of constraints,
let us note that the operator
$L_0$ applied to the state \p{Fockvector} acts
as a usual covariant d'Alambertian
$\nabla_\mu \nabla^\mu $ on the
tensor $\Phi_{\alpha_1\alpha_2\cdots\alpha_{n}}
(x)$. As we shall see below, the ``actual" mass -shell operator, which
annihilates the physical state, can be constructed on
the basis of the
above operator $L_0$ (see eqs. \p{ln} and \p{ms}).}
\be \label{lapl}
L_0 = g^{\mu
\nu}(p_\mu p_\nu - \Gamma^\lambda_{\mu \nu}p_\lambda)= p^\alpha
p_\alpha + \omega_\alpha{}^{\alpha \beta} p_\beta \ee where the last
equality in \p{lapl} is due to the torsionlessness of the $AdS_D$ metric.
Using \p{osc} and the identities \begin{equation} [p_\mu,p_\nu]=
-r(a_\mu^+ a_\nu - a_{\nu}^+ a_{\mu}), \end{equation} \begin{equation}
[p_\mu, a^{\nu +}] = - \Gamma^\nu_{\mu \lambda}a^{\lambda +}, \quad
[p_\mu, a^{\nu}] = - \Gamma^\nu_{\mu \lambda}a^{\lambda} \end{equation}
with the parameter $r$ being related to the inverse radius of
$AdS_D$ space $\lambda$ via $r = \lambda^2$, one can straightforwardly
 check that the  commutation relations among the operators $L_1^\pm,
L_2^\pm,$ and the modified d'Alambertian
\begin{equation} \label{ln}
\tilde L_0 \equiv L_0 + r(-D + \frac{D^2}{4} + 4L^+_2  L_2 - G_0
G_0 + 2G_0),
\end{equation}
where $G_0 = a^{\mu+} a_{\mu} +\frac{D}{2}$,
form the closed algebra \cite{BMV}
\begin{eqnarray} \label{al}
&&[L^+_1, L_1] = -\tilde L_0,  \quad
[L^+_1, L_2] = -L_1, \quad
[L^+_2, L_1] = -L^+_1, \nonumber \\
&&[\tilde L_0, L_1] = -2rL_1 + 4rG_0  L_1 -
      8rL^+_1  L_2, \nonumber\\
&&[L^+_1, \tilde L_0] = -2rL^+_1 + 4rL^+_1  G_0 -
      8rL^+_2  L_1,  \nonumber \\
&&[G_0, L_1] = -L_1, \quad [L^+_1, G_0] = -L^+_1,
\end{eqnarray}
with the $SO(2,1)$ subalgebra
\begin{equation} \label{so21}
[G_0, L_2] = -2L_2, \quad
[L^+_2, G_0] = -2L^+_2, \quad
[L^+_2, L_2] = -G_0.
\end{equation}
After the modification of the d'Alambertian given by \p{ln}
 the nonlinear algebra \p{al} can be considered as  deformation
of the corresponding Lie algebra for the flat -- space case with
the  parameter
$r$ being a deformation parameter. Note the proportionality
of the last term in \p{ln} to the Kasimir operator of the
$SO(2,1)$ linear subalgebra \p{so21}.
This modification of the d'Alambertian leads
 to a correct physical spectrum,
 as we shall see below.

The next step is to construct the nilpotent BRST charge, corresponding
to the system \p{al} -- \p{so21}, in order to use the method of
construction of BRST invariant Lagrangians \cite{OS} -- \cite{BPT2},
which automatically leads to the proper equations to be
satisfied by  physical fields
 along with the required gauge invariance needed to remove the
nonphysical states from the spectrum.
However,
the problem of constructing  the corresponding nilpotent BRST
-charge for this system
of operators under consideration is twofold. First, as it was in the
case of  flat space -- time, the operator $G_0$, which is the Cartan
generator of the algebra \p{so21}, is strictly positive and it cannot
annihilate any nonzero state in the Fock space and, therefore, has to be
excluded from the total set of constraints.  This means that we have
 second class constraints in our system and, therefore, the standard
method of BRST construction does not work.  The general prescription
for finding the nilpotent BRST charges for the system of
constraints, which form an arbitrary Lie algebra, but an arbitrary number
of the Cartan generators are excluded from the total set of constraints
due to some physical reasons, was given in \cite{BPT1}.  The essential
point of this method is that though actually one deals with the
system of  second class constraints, before exclusion of the Cartan
generators, one  formally has
the system of  first class constraints,
and the corresponding ``bare" BRST charge
can be written straightforwardly. Further, the ``actual" BRST charge
can be obtained from the ``bare" one after  special similarity
transformations.

The second problem arising when we investigate the system
\p{al} -- \p{so21}
is that  the operators under consideration form the nonlinear algebra.
Note also that for  higher spin fields propagating in  $AdS_D$ space
it is impossible to omit the $L_2^\pm$ operators, thus simplifying the system,
and
study just reducible representations of the $AdS_D$ group, since
the operators $L_2^\pm$ appear automatically in the right -- hand side
of the commutators containing the
operators $L_1^\pm$.

The nonlinear character of  algebra \p{al} inherent in $AdS$ space
 is a new aspect in
deriving  higher spin Lagrangians within the framework of
the  BRST
-- method. The structure of the BRST charge, especially the form of its
terms trilinear in ghosts, is analogous
to the construction for nonlinear $W$ -- algebras \cite{TM}.

Below, we construct the nilpotent BRST charge
for the system \p{al} -- \p{so21}
appropriately generalizing the method of such constructions
developed for an arbitrary Lie algebra (\cite{BPT2}, \cite{BPT1}).
Let us first introduce the anticommuting ghost variables
$\e_0, \e^+_1, \e_1, \e^+_2, \e_2, \e_G$ which correspond to the operators
$\tilde L_0, L_1, L^+_1, L_2, L_2^+$ and $G_0$ and have the ghost number equal to $1$,
as well as the corresponding momenta
$\cp_0, \cp_1, \cp^+_1, \cp_2, \cp^+_2, \cp_G$ with the ghost number $-1$
obeying the anticommutation relations
\begin{equation}
\{\e_0,\cp_0\}=\{\e_2,\cp_2^+\}=\{\e_2^+,\cp_2\}=\{\e_2,\cp_2^+\}=\{\e_2^+,\cp_2\}=
\{\e_G,\cp_G\}=1.
\end{equation}
Then we build the auxiliary representations of the generators of the $SO(2,1)$
algebra \p{so21} in terms of the additional creation and annihilation operators
$b^+$ and $b$, $[b, b^+]=1$ in terms of the Verma module.
Namely, introduce the vector in the space of  Verma module
$|n\rangle = {(L_2^+)}^n|0\rangle_V$, $n \in N,$ $L_2|0\rangle_V=0$
and the corresponding vector in the Fock space
$$
|n \rangle = {(b^+)}^n|0\rangle, \quad b|0 \rangle =0.
$$
Mapping the vector in the space of Verma module to the vector in the Fock space one obtains
the representations of the generators of algebra $SO(2,1)$ in terms of the
variables $b^+$ and $b$
\begin{equation} \label{aux}
L^+_{2.aux} = b^+, \quad
G_{0.aux} = 2b^+  b + h, \quad
L_{2.aux} = (b^+  b + h)  b
\end{equation}
where $h$ is a parameter.
Then defining, the operators $\tilde L^\pm_2 = L^\pm_2 + L^\pm_{2.aux},
\tilde G_0 = G_0 + G_{0.aux}$, we construct
the corresponding ``bare" nilpotent BRST charge for the
total system of constraints
\p{al}--\p{so21} adapting the procedure developed for nonlinear
$W$ - algebras in \cite{TM}
\begin{eqnarray} \label {brst1}
Q_1\!\!& = &\!\!\!\e_0(\tilde L_0- 4 r G_{0.aux}+6r)\!+ \!\e_1 L^+_1 \! +
     \! \e^+_1  L_1\! -\!\e_2 (L^+_2 + L^+_{2.aux})\!\! -
\!\! \e^+_2(L_2 +L_{2.aux})\!\! +
     \nonumber \\
      &&\!\! \e_G (G_0 + 2b^+b+h -3+
     \e^+_1  \cp_1  + \cp^+_1  \e_1+2 \cp^+_2  \e_2+2 \e^+_2   \cp_2)- \nonumber \\
     &&\!\! (\e^+_2  \e_2 +4 r \e^+_1  \e_1  +
     8 r \e_0  \e^+_1  \cp^+_1  \e_2+
     8 r \e_0  \e^+_2  \e_1  \cp_1 ) \cp_G-\nonumber \\
      &&\!\!  \e^+_1  \e_1  \cp - \e^+_1  \cp^+_1  \e_2+
     \e^+_2  \e_1  \cp_1+2 r \e_0  \e^+_1  \cp_1 + 2 r \e_0  \cp^+_1  \e_1-
     \nonumber \\
      &&\!\!\!\!\!4 r \e_0 ( \e^+_1 \cp_1 G_0\! +\! 2  \e^+_1  \cp_2 L^+_1\!
     +\!2  \cp^+_2 \e_1 L_1\!+ \! \cp^+_1 \e_1 G_0\!+
\!2 \e^+_1  \cp^+_1  L_{2.aux}\! -\!
    2\e_1  \cp_1 L^+_{2.aux})\!\!+ \nonumber \\
     &&\!\!\!\!\!  4 r \e_0(2\e^+_1  \e^+_2  \cp_1  \cp_2\! +\!
    \e^+_1  \cp^+_1  \e_1  \cp_1\! -\!2\e^+_1  \cp^+_2  \e_2  \cp_1 \! -
    \!2\e^+_2  \cp^+_1  \e_1  \cp_2\! +\!
    2\cp^+_1  \cp^+_2  \e_1  \e_2). \nonumber
\end{eqnarray}
In order to avoid the condition of the type $G_0| \Phi \rangle =0$
one has to get rid of the variables $\e_G$ and $\cp_G$ in the BRST charge
\p{brst1}, keeping its nilpotency property at the same time.
This can be  done by replacing the parameter $h$ by the expression
$-(G_0 + 2b^+b -3+
     \e^+_1  \cp_1  + \cp^+_1  \e_1+2 \cp^+_2  \e_2+2 \e^+_2   \cp_2) $
and then simply omitting the dependence on  $\cp_G$ in the BRST
 charge \cite{BPT1}.
 The ``reduced" BRST charge thus takes the form
\begin{eqnarray} \label{brst}
Q& =&\!\!\e_0(\tilde L_0 + 4 r G_0 - 6 r)\!
  -\! \e_2(L^+_2 + b^+)\!
       +\!\e_1L^+_1\!  + \!\e^+_1  L_1\! -
    \!  \e^+_2 ( L_2- G_0b + b - b^+bb)\!- \nonumber \\
       &&
      \e^+_1  \e_1  \cp
      - \e^+_1  \cp^+_1  \e_2  +
      \e^+_2  \e_1  \cp_1+
      2r \e_0(3\e^+_1  \cp_1 + 4 \e^+_2  \cp_2
      +3 \cp^+_1  \e_1 + 4 \cp^+_2  \e_2)+ \nonumber \\
       && 8 r \e_0    \e_1  \cp_1 b^+ -
      4 r \e_0  \e^+_1    \cp_1 G_0
      - 8 r \e_0  \e^+_1   \cp_2 L^+_1 +
      8 r \e_0  \e^+_1  \cp^+_1(G_0 b -  b)- \nonumber \\
      && 4 r \e_0  \cp^+_1   \e_1 G_0-
      8 r \e_0  \cp^+_2    \e_1 L_1 - \e^+_1  \e^+_2    \cp_1 b
       + \e^+_2  \cp^+_1    \e_1 b +
      2 \e^+_2  \cp^+_2    \e_2 b +\nonumber \\
      &&  8 r \e_0  \e^+_1  \e^+_2  \cp_1  \cp_2
       +
      4 r \e_0  \e^+_1  \cp^+_1  \e_1  \cp_1
      -8 r \e_0  \e^+_1  \cp^+_2  \e_2  \cp_1- \nonumber \\
      &&8 r \e_0  \e^+_2  \cp^+_1  \e_1  \cp_2
      +8 r \e_0  \cp^+_1  \cp^+_2  \e_1  \e_2 -
      16 r \e_0  \e^+_1  \e^+_2  \cp^+_1    \cp_2 b+ \nonumber \\
      &&  8 r \e_0  \e^+_1  \cp^+_1  b^+  b  b
      +
      16 r \e_0  \e^+_1  \cp^+_1  \cp^+_2    \e_2 b.
\end{eqnarray}
Finally, to restore the hermiticity property of the BRST charge, which
is lost due to the form of auxiliary representations \p{aux},
one has to define the scalar product in the Fock space as
$\langle \Phi_1|K|\Phi_2 \rangle$ were $K$ is a nondegenerate kernel operator
\be \label{yadro}
K= Z^+Z, \quad Z=\sum_{n} \frac{1}{n!}
 {(L_2^+)}^n |0 \rangle_V \langle 0|{(b)}^n
\ee
satisfying the property $KQ=Q^+K$. Note also that $K=1$ in the $b^+$
 independent sector of the Fock space.
Thus,  the construction of the nilpotent BRST charge
with an appropriate hermiticity condition is fully completed.

\setcounter{equation}0\section{The Lagrangian}

The procedure of constructing  the BRST invariant Lagrangian
is realized analogously to
 \cite{OS} -- \cite{BPT2}.
After defining the ghost vacuum
\be
\e_1|0\rangle_{gh}=\e_2|0\rangle_{gh}=\cp_1|0\rangle_{gh}=
\cp_2|0\rangle_{gh}=\cp|0\rangle_{gh}=0,
\ee
we can write the BRST invariant Lagrangian in the form
\begin{equation} \label{L}
{\cal L} = \int d \e_0 \langle \chi| K Q | \chi \rangle,
\end{equation}
By the construction this Lagrangian is invariant under the gauge
transformations \begin{equation} \label{G} \delta | \chi \rangle = Q |
\Lambda \rangle \end{equation} with a vector $| \Lambda \rangle$
treated as a parameter of gauge transformations. The state vector
$|\chi \rangle$ and a  parameter of gauge transformations $|\Lambda
\rangle$ have the ghost numbers $0$ and $-1$, respectively, and, therefore,
can be written as follows:
 \begin{eqnarray} \label{vector}
|\chi\rangle&=&|S_0\rangle + {\eta}_1^+ {\cal P}_1^+ |S_1\rangle +
{\eta}_2^+ {\cal P}_1^+ |S_2\rangle + {\eta}_1^+ {\cal P}_2^+
|S_3\rangle +{\eta}_2^+ {\cal P}_2^+ |S_4\rangle + {\eta}_1^+
{\eta}_2^+ {\cal P}_1^+ {\cal P}_2^+ |S_5\rangle \nonumber \\ &&+\eta_0
{\cal P}_1^+ |R_1\rangle +\eta_0 {\cal P}_2^+ |R_2\rangle + \eta_0
{\eta}_1^+ {\cal P}_1^+ {\cal P}_2^+ |R_3\rangle + \eta_0 {\eta}_2^+
{\cal P}_1^+ {\cal P}_2^+ |R_4\rangle \end{eqnarray}

\begin{eqnarray}\label{lambda}
|\Lambda\rangle &=&
{\cal P}_1^+ |\lambda_1\rangle +
{\cal P}_2^+ |\lambda_2\rangle +
{\eta}_1^+ {\cal P}_1^+ {\cal P}_2^+ |\lambda_3\rangle +
{\eta}_2^+ {\cal P}_1^+ {\cal P}_2^+ |\lambda_4\rangle +
\eta_0 {\cal P}_1^+ {\cal P}_2^+ |\lambda_5\rangle
\end{eqnarray}
with the vectors $|S_i \rangle$, $|R_i \rangle$ and $|\lambda_i \rangle$
having ghost number zero and depending only
on bosonic creation operators $a_\mu^{+}, b^+$.

Before proceeding further let us note that
in the absence of the
$b^+$ dependence and  other fields in  expression \p{vector},
the BRST charge \p{brst}, acting on the $\ra{0}$ component
of the state vector $|\chi \rangle$,
forms the mass -- shell condition
\begin{equation} \label{ms}
(\tilde L_0 + 4rG_0  -6r)\ra{0}=0
\end{equation}
along with the transversality and tracelessness conditions \p{nogh}, \p{notr}.
This means that if one excludes all auxiliary fields such as
 $\ra{i}, (i=1,...,5)$ and
$|R_i \rangle, (i=1,...,4)$ and the $b^+$ dependence of $\ra{0}$
using the BRST gauge invariance
with the parameters $|\lambda_i\rangle$ ($i=1.,,,5$),
the physical field $\ra{0}$ being transversal
and traceless, will satisfy the proper mass -- shell
condition obtained in \cite{Me}.
Therefore, the ``actual" mass -- shell conditions (3.6) obtained from the
requirement of nilpotency of the BRST charge for the system \p{al}
-- \p{so21} completely coincides with the one derived from the analysis
of the unitary representations of the $AdS_D$ group \cite{Me}. At this
point one should note that the inclusion of any additional terms in the
mass-shell condition \p{ms} will inevitably  destroy
the nilpotency of the BRST charge.  The restoration of this crucial
property will lead to the necessity of introducing  additional
creation operators in the considered Fock space (see, for example,
\cite{PT1}), thus explaining the
appearance of additional fields in the Lagrangians for {\em massive}
higher spins.

Writing explicitly the equations of motion following
from the Lagrangian \p{L} and the BRST gauge transformations \p{G}
in terms of the component fields $\ra{i}, |A_i\rangle$ and
$|\lambda_i\rangle$, one can prove in  complete analogy with \cite{PT2}
that all fields except the $b^+$ independent part of $\ra{0}$ can be
gauged away, the last one being subject to the conditions \p{nogh},
\p{notr}, \p{ms}. As we have mentioned above,  the latter fact guarantees that the
Lagrangian \p{L} correctly describes the unitary representations of
the $AdS_D$ group \cite{Me}.

On the other hand, it is possible to gauge away only
 part of the auxiliary fields
using the BRST gauge invariance, thus not fixing completely
the values of gauge parameters $| \lambda_i \rangle$.
 The rest of the fields can be expressed
via the basic field $\ra{0}$ and inserted back into the Lagrangian
\p{L}. Since we do not fix
the gauge parameters completely, the Lagrangian will possess
some residual gauge invariance. As was shown in \cite{BPT2},
in the case of Minkowski flat space such a line of construction
leads straightforwardly to the Fronsdal Lagrangian \cite{F}.

Indeed one can prove, following
the lines of \cite{BPT2} (see the Appendix),
that the vector $| \chi \rangle$ contains the only physical
field $| S_0 \rangle$ with no $b^+$ dependence and according
to our previous discussion,
it will be the basic field $|\Phi \rangle$
(i.e. $\ra{0} \equiv |\Phi \rangle$).
 The other fields
can be either excluded using the equations of motion or gauged away.
Then, after the partial BRST gauge fixing
we arrive at the  Lagrangian describing the irreducible massless
integer higher spin fields interacting with the $AdS_D$ background
\begin{eqnarray} \label{LF}
\cal L&=&\langle \Phi|
 \tilde L_0 - L_1^+L_1 - 2 L^+_2  \tilde L_0  L_2
 +L^+_2  L_1  L_1
     +L^+_1  L^+_1  L_2   - L^+_2 L^+_1 L_1 L_2 \nonumber \\
 &&- r(6  - 4    G_0 + 10    L^+_2  L_2      -
    4  L^+_2  G_0  L_2)
|\Phi \rangle
\end{eqnarray}
where due to the auxiliary conditions
imposed on the fields $|\chi\rangle$
or, equivalently due to the equations of motion \p{S1} -- \p{S2}
with respect to the auxiliary fields, the basic field $|\Phi \rangle$
 is double traceless
\be
L_2L_2|\Phi \rangle=0.
\ee

The Lagrangian \p{LF}, as a consequence of equations \p{G} and \p{gp}, is
invariant under the transformations \be \label{k} \delta |\Phi \rangle
= L^+_1|\lambda_1 \rangle \ee with the traceless parameter of gauge
transformations $L_2|\lambda_1 \rangle=0$.  As it is shown in the
appendix the tracelessness of the parameter of the gauge transformation
follows after gauging away  the auxiliary field $\ra{2}$,
thus restricting the remaining gauge freedom.

The Lagrangian \p{LF} and the gauge transformations can be written
 in a more conventional tensorial notation.  Namely, after introducing the usual
 covariant derivative $\nabla_\mu$ and using the expression \p{Fockvector}
 for the double traceless
symmetric field $\Phi_{\mu_1 \mu_2 \cdots \mu_s}(x)$, having the spin $s$,
we obtain the action
\begin{eqnarray} \label{action}
S&=& \int d^Dx \sqrt{-g} (\Phi^{\mu_1 \mu_2 \cdots \mu_s}(x)
(\nabla^2 - r(s^2 + sD - 6s - 2D +6))
\Phi_{\mu_1 \mu_2 \cdots \mu_s}(x) \nonumber \\
&& - \frac{s(s-1)}{2}
\Phi_{\mu_1}{}^{\mu_1 \mu_3 \cdots \mu_s}(x)
(\nabla^2 - r(s^2 +sD - 4s - D +1))
\Phi^{\nu_1}{}_{\nu_1 \mu_3 \cdots \mu_s}(x)  \nonumber \\
&&-s \Phi^{\mu_1 \mu_2 \cdots \mu_s}(x) \nabla_{\mu_1} \nabla_{\nu_1}
\Phi^{\nu_1}{}_{\mu_2 \cdots \mu_s}(x) \nonumber \\
&&+s(s-1) \Phi_{\mu_1}{}^{\mu_1 \mu_3  \cdots \mu_s}(x) \nabla_{\nu_1} \nabla_{\nu_2}
\Phi^{ \nu_1 \nu_2}{}_{\mu_3 \cdots \mu_s}(x) \nonumber \\
&&-\frac{s(s-1)(s-2)}{4}
\Phi_{\mu_1}{}^{\mu_1 \mu_3 \mu_4 \cdots \mu_s}(x) \nabla_{\mu_3} \nabla_{\nu_1}
\Phi^{\nu_1 \nu_2}{}_{\nu_2 \mu_4 \cdots \mu_s}(x))
\end{eqnarray}
which is invariant under the gauge transformations
\be \delta \Phi_{\mu_1 \mu_2 \cdots  \mu_s}(x) = \nabla_{\{ \mu_s}
\lambda_{\mu_1 \cdots \mu_{s-1} \}}(x)
\ee
with the totally symmetric and traceless
($\lambda^{\mu_1}{}_{\mu_1 \mu_3 \cdots ... \mu_{s-1}}(x)=0$)
parameter of gauge transformations. Eq (3.10) is our final result.

The action \p{action}  contains the nonminimal interaction of  higher spin fields
with the gravitational field and coincides
with the one given in \cite{F1}  for the case of $D=4$
(see also \cite{FV} for the equivalent action
written with the use of a different formalism)
while for arbitrary $D$ dimensions it gives the part (depending on the
higher rank field) of the massive theory action
of \cite{Z}
in which the mass parameter is set to be equal to zero. To describe
the massless representations, we should consider
the doubly traceless fields. This very important property of the
basic field does not take place at the formal massless limit of massive
theory formulated in \cite{Z}.

On the other hand, in the flat space - time  limit,
i.e., when the parameter $r$ is set equal to zero, one recovers the
$D$ dimensional generalization of the Fronsdal Lagrangian \cite{F}
describing free  massless fields with a integer spins.
Obviously, the latter Lagrangian can  also be constructed
following the procedure given in the paper \cite{PT2}
if we put initially  $r=0$
in the algebra of constraints \p{al} -- \p{so21}.

\setcounter{equation}0\section{Conclusions}
In this paper, we have constructed the field theoretical Lagrangian
for the massless irreducible higher integer spin fields propagating
through the AdS background with an arbitrary dimension $D$.
For this purpose,
we have used the method of BRST constructions
adopted for the system of the
arised here second class constraints and nonlinear algebras.

The subsequent procedure of construction of the corresponding BRST
invariant goes in complete analogy with the one developed in
 \cite{PT1} -- \cite{BPT2} for the description of the free massive and
 massless fields belonging to various representations of the Poincare
group.  Due to this analogy the generalization of the approach under
 consideration to  more complicated representations of the $AdS_D$
 group \cite{BMV} seems to be quite possible. Another interesting
 problem is the application of this approach to the description of the
 massive higher spin fields in the $AdS_D$ background \cite{BGP} --
\cite{KL}, thus extending the
 results of \cite{PT1} obtained for  flat
space-time.  Namely, modifying the d'Alambertian \p{lapl} by the terms
proportional to mass and scalar curvature $R$, one arrives at the
 massive higher spin fields with general nonminimal coupling to
 gravity.  This changes the system \p{al} -- \p{so21} to the one which
 contains the second class constraints even before the exclusion of the
 operator $G_0$, and the construction of the
 corresponding nilpotent BRST
 charge can be more complicated.

We hope to touch upon these issues as well as to analyze the
applications of the BRST construction to  arbitrary higher spin
 fermionic fields on  $AdS_D$ space and  investigating the
 possibilities of developing the  theory on the gravitational backgrounds with
 nonconstant curvature in the subsequent publications.  \vspace{0.3cm}

\noindent {\bf Acknowledgments}
We are grateful to Prof. M.A. Vasiliev for the interest to the work
and valuable comments. A.P would like to thank Prof. O. Lechtenfeld
and Prof. N. Dragon for kind hospitality at the Institute for
Theoretical Physics, University of Hannover where part
of this work has been done.
The work of the authors was supported in part by
 INTAS grant, project No 00-00254. The work of I.L.B and A.P.
was also supported in part by the joint DFG-RFBR grant, project No
99-02-04022. I.L.B is grateful to  RFBR grant, project No
99-02-16617 and to  INTAS grant, project No 991-590 for partial
support. The work of A.P and M.T was supported in part by  RFBR
grant, project No 99-02-18417. M.T is grateful to  RFBR grant,
project No 01-02-06531 for partial support.

\vspace{1cm}
\setcounter{equation}0
\section*{ Appendix }
\def\theequation{A\arabic{equation}}

\vspace{.5cm}
In order to explicitly derive the Lagrangian \p{LF},
 let us rewrite  the
vector  $| \chi \rangle$    and the parameter $| \Lambda \rangle$
in the form
\begin{equation} \label{expans}
| \chi \rangle = | \chi_0 \rangle + \cp_2^+| \chi_2 \rangle, \quad
| \Lambda \rangle = | \Lambda_0 \rangle + \cp_2^+|
\Lambda_2\rangle
\end{equation}
From the gauge transformation law \p{G} and the explicit form of the
auxiliary representations of $SO(2,1)$ algebra \p{aux} one can see
that the field  $| \chi_0 \rangle$    transforms through the parameter
$|\Lambda_2 \rangle$
as
\begin{equation}
\delta| \chi_0 \rangle = (L^{+}_2 + b^+)| \Lambda_{2} \rangle
\end{equation}
and, therefore, the parameter $| \Lambda_{2} \rangle$
can be used to remove the $b^+$ dependence in the vector
 $| \chi_0 \rangle$,  i.e.,
\begin{equation} \label{NB}
b| \chi_0 \rangle = 0
\end{equation}
at the same time one is left with the gauge freedom with the parameter
$|\Lambda_0 \rangle$ which is now also independent of $b^+$.

Next, we impose the auxiliary conditions on the vector $|\chi \rangle$.
Namely, since the Kernel operator \p{yadro}
is nondegenerate and, therefore, has the inverse,
we obtain from the Lagrangian
\p{L} the following equations of motion:
\be \label{UD}
Q|\chi \rangle =0.
\ee
We can also impose the conditions
\be
B_I|\chi \rangle = [M_I, Q]_\pm |\chi \rangle =0
\ee
for some operators $M_I$
without affecting the physical content of the theory \cite{M}.
Obviously,  these auxiliary conditions remove zero norm states being
the ``pure BRST gauge".
The choice of operators $M_I$ is actually the choice of BRST gauge.
Taking $M_1 =b$, we obtain the equation
\begin{equation}
(8 r \e_0  \e_1  \cp_1 + \e^+_2  b  b +
  8 r \e_0  \e^+_1  \cp^+_1  b  b - \e_2)|\chi \rangle =0
\end{equation}
which with regard to  equation \p{NB},
can be solved as
$|\chi_2 \rangle = 8r \e \ra{1}.$
Therefore, the state vector $\chi$ has now the following reduced form:
\be \label{vn}
|\chi \rangle = \ra{0} + (\e_1^+ \cp_1^+ -8r \e \cp_2^+)\ra{1}
+\e_2^+ \cp_1^+ \ra{2}  + \e \cp_1^+ |R_1 \rangle.
\ee
Imposing one more auxiliary condition on $|\chi \rangle$
with $M_2 = \cp_2$, we get the set of equations on the fields left in \p{vn}.
\be \label{S1}
\ra{1} = L_2\ra{0}
\ee
and
\be \label{S2}
L_2 |R_1 \rangle =L_2\ra{2} = L_2 \ra{1}=0
\ee
The remaining fields $\ra{0}, \ra{1}, \ra{2}$ and $|R_1 \rangle$
transform under the residual BRST gauge transformations as follows:
\begin{equation} \label{g1}
\delta \ra{0} = L_1^+ |\lambda_1 \rangle, \quad
\delta \ra{1} = L_1 |\lambda_1 \rangle,
\quad
\delta |R_1\rangle = \tilde L_0|\lambda_1 \rangle
\end{equation}
and
\begin{equation} \label{g2}
\delta \ra{2} = -L_2 |\lambda_1 \rangle
\end{equation}
with unconstrained parameter  $|\lambda_1 \rangle$.
From the gauge transformation law \p{g2} one can see
that the field $\ra{2}$ can be gauged  away
 and one is left  with the residual gauge invariance with the traceless
parameter
\be \label{gp}
L_2|\lambda_1 \rangle=0.
\ee
Inserting the state vector \p{vn} with $\ra{2}=0$ into the Lagrangian \p{L}
and performing the integration with respect to the ghost variables,
we arrive at
\begin{eqnarray} \label{L1}
\cal L&=&\langle R_1|  |R_1 \rangle  +
  \langle R_1|  L^+_1  \ra{1} +
\la{1}  L_1  |R_1 \rangle -  \la{1} 8r L_2  \ra{0}- \la{0}8r  L^+_2  \ra{1}
\nonumber \\
  && +\la{0}  4rG_0 - 6r +\tilde L_0 \ra{0}+
     \la{1}  4rG_0 + 6r -\tilde L_0  \ra{1}  \nonumber \\
&&-\la{0}  L^+_1  |R_1 \rangle - \langle R_1|  L_1  \ra{0}.
   \end{eqnarray}
  One can easily check that all equations of motion
resulting from the Lagrangian \p{L},
 except the ones  resulting after the variation with
respect to the fields $\ra{0}, \ra{1}$ and $|R_1 \rangle$, coincide with
\p{S1}--\p{S2} and, therefore, no additional conditions on these fields are lost.
Finally, expressing the field $|R_1 \rangle$ due to its own equation of motion
$L^+_1  \ra{1} - L_1  \ra{0} + |R_1 \rangle=0$
and  inserting
back into  \p{L1} along with  expression \p{S1}
we arrive at \p{LF} for $\ra{0} \equiv |\Phi \rangle$.
As a consequence of \p{S1} and last equation in \p{S2}, the field
$\ra{0}$ is doubly traceless $L_2L_2\ra{0}=0$.

\end{document}